\documentclass[12pt,a4paper]{article}
\usepackage[a4paper,top=3cm,bottom=3cm,left=2cm,right=3cm,bindingoffset=5mm]{geometry}
\usepackage[bottom]{footmisc}

\usepackage[utf8]{inputenc}
\usepackage{jheppub}
\usepackage{xcolor}

\usepackage{tikz}

\usepackage{amsmath}
\usepackage{amsthm}
\usepackage{amssymb}
\usepackage{physics}
\usepackage{comment}
\usepackage{graphicx}
\usepackage{float}
\usepackage{hyperref}
\usepackage{amsfonts}
\usepackage{mathrsfs}
\usepackage{xcolor}

\usepackage{adjustbox}

\usepackage{scalerel}

\hypersetup{
colorlinks=true,         
linkcolor=blue,          
citecolor=red,        
urlcolor=blue            
}

\newcommand{\bz}{{\bar{z}}}
\newcommand{\tlambda}{\tilde{\lambda}}
\newcommand{\bT}{\mathbf{T}}

\newcommand{\lr}[1]{\langle #1 \rangle}

\DeclareFontFamily{U}{jkpmia}{}
\DeclareFontShape{U}{jkpmia}{m}{it}{<->s*jkpmia}{}
\DeclareFontShape{U}{jkpmia}{bx}{it}{<->s*jkpbmia}{}
\DeclareMathAlphabet{\mathfrakalt}{U}{jkpmia}{m}{it}
\SetMathAlphabet{\mathfrakalt}{bold}{U}{jkpmia}{bx}{it}

\newcommand{\myw}{\mathfrakalt{w}}

\newcommand{\Lw}{\text{L}\myw_{1+\infty}}

\title{Generating Hodges' Graviton MHV Formula  with an $\Lw$ Ward Identity}
\author[1]{Alfredo Guevara,}
\author[2,3]{Elizabeth Himwich,}
\author[1,3]{and Noah Miller}

\affiliation[1]{School of Natural Sciences, Institute for Advanced Study}
\affiliation[2]{Princeton Center for Theoretical Science, Princeton University}
\affiliation[3]{Princeton Gravity Initiative, Princeton University}

\emailAdd{aguevara@ias.edu}
\emailAdd{ehimwich@princeton.edu}
\emailAdd{noahmiller@ias.edu}

\abstract{Hodges' formula expresses the tree-level all-multiplicity Einstein gravity MHV amplitude as a matrix determinant. In this work, we  prove that Hodges' determinant is generated by an $\Lw$ Ward identity on the celestial sphere. The Ward identity takes the form of a recursion relation that has not previously appeared in the literature and is unrelated to BCFW. The proof makes use of the matrix-tree theorem.

}

\begin{document}

\maketitle

\section{Introduction}

Is there a holographic dual to the gravitational $S$-matrix in asymptotically flat spacetime? Success in reformulating perturbative gauge theory (with and without supersymmetry) using methods agnostic to Feynman diagrams have provided strong encouragement that similar constructions may exist in gravity. The dual descriptions of gauge theory not only streamline cumbersome calculations, but also pave the way to taming divergences of the theory \cite{Arkani-Hamed:2010zjl,Arkani-Hamed:2010pyv,Arkani-Hamed:2023epq}. 

A great deal of this progress has been based on the use of extended symmetries, such as those arising from integrability \cite{Bena:2003wd,Dolan:2004ps,Alday:2007hr,Drummond:2008vq,Gromov:2009tv,Gromov:2009bc}, which for instance played a pivotal role in the reformulation of super-Yang-Mills scattering amplitudes with twistor variables. This reformulation put forth a new paradigm by suggesting that perhaps the actual scattering fundamentally occurs not in spacetime, but in a dual space \cite{Witten:2003nn,Roiban:2004yf,Hodges:2009hk,Arkani-Hamed:2009hub,Arkani-Hamed:2009ljj,Mason:2013sva}. In the dual description, locality and unitarity emerge  in a completely novel way and the existence of spacetime Feynman diagrams appears fortuitous.

In pursuit of an analogous construction for gravity, Hodges introduced new dual representations of gravitational amplitudes in a series of breakthrough papers culminating in the celebrated formula for tree-level MHV scattering \cite{Hodges:2005bf,Hodges:2005aj,Hodges:2009hk,Hodges:2011wm,Hodges:2012ym}. The construction was guided by the appearance of momentum twistors and a specific theory of gravity with $\mathcal{N}=7$ supersymmetry, as well as by the NSVW tree graph formula \cite{Nguyen:2009jk,Bern:1998sv}. It has generated important progress in many directions \cite{Cachazo:2012kg,Cachazo:2012pz,Cachazo:2013iea,Herrmann:2016qea} but remains mysterious despite much effort to be generalized.

Jumping forward to today, new tools from the celestial holography program are providing deep hints about the structure of gravitational scattering. In particular, a new extended $\Lw$ symmetry has been identified in the tree-level amplitudes of minimally-coupled Einstein gravity \cite{Guevara:2021abz,Strominger:2021mtt}. In analogy with gauge theory, this suggests that a dual description for gravitational scattering could exist in a space that makes the extended $\Lw$ symmetry manifest, rather than spacetime. Coincidentally, in analogy with the integrable symmetries of gauge theory, the $\Lw$ symmetry is again realized in twistor space, though it  can also be pulled down to a holographic description on the two-dimensional (2D) celestial sphere \cite{Adamo:2021lrv,Adamo:2021zpw,Costello:2022wso,Guevara:2022qnm,Costello:2023hmi}.

The discovery of $\Lw$ symmetry within gravitational scattering immediately raises the question of how it can be used concretely to constrain and construct amplitudes, particularly from the celestial dual perspective \cite{Mago:2021wje,Freidel:2021ytz,Ren:2022sws,Banerjee:2023zip}. From the spacetime perspective, $\Lw$ symmetry organizes a universal part of the soft expansion of any tree-level minimally-coupled gravitational scattering amplitude \cite{Guevara:2019ypd,Himwich:2021dau,Guevara:2022qnm,Himwich:2023njb}, which raises another  question of whether the universal part  corresponds exactly to a class of known amplitudes. These questions both touch on whether a particular known set of gravitational amplitudes is specified fully by (dual) $\Lw$ symmetry. This was asked specifically about Hodges' formula in \cite{Guevara:2022qnm,Adamo:2021bej} where a positive partial answer was provided.\footnote{See \cite{Banerjee:2020zlg,Banerjee:2021cly,Banerjee:2021dlm,Ren:2023trv} for related work on  celestial OPEs in gravity MHV amplitudes and \cite{Mol:2024etg} for an alterative dual construction of the graviton MHV leaf correlator.}

Taking a step in this direction, we demonstrate that Hodges' formula for tree-level MHV gravitational scattering amplitudes is exactly generated by a novel recursion relation that corresponds to the 2D Ward identity of $\Lw$ symmetry in the dual celestial description.\footnote{From the realization of $\Lw$ in twistor space, one might naturally expect that the amplitudes generated by $\Lw$ would be those of the self-dual sector, although self-dual amplitudes vanish at tree level. There may however be an analogous formula for the one-loop amplitude, where it is known that the $\Lw$ symmetry is perturbatively exact \cite{Ball:2021tmb,Bittleston:2022jeq}. We hope to explore this in future work.} 
We believe it provides strong evidence that a dual of the gravitational $S$-matrix exists on the 2D celestial sphere. Related work on the gauge theory MHV amplitude has appeared in \cite{Parke:1986gb,Nair:1988bq,Adamo:2021zpw,Bu:2022dis,Adamo:2022wjo,Costello:2022wso,Bu:2023vjt,Melton:2024akx,Bittleston:2024efo,Donnay:2025yoy}.

\subsection{Main Result}

Consider a tree-level maximally-helicity-violating (MHV) scattering amplitude $\mathcal{M}_n$ in pure Einstein gravity, with external gravitons $1^{--}, 2^{--}, 3^{++}, \ldots, n^{++}$,
\begin{equation}\label{eq:strippedamplitude}
    \mathcal{M}_n =  (2 \pi)^4 \delta^4( p_1 + \ldots + p_n) M_n
\end{equation}
where $M_n$ is the stripped amplitude. Crucially, $M_n$ is regarded as a function of all the momenta $\{ p_i \}$ without the constraint of momentum conservation. Thus it is only defined up to a function that vanishes on this constraint. However, Hodges \cite{Hodges:2012ym} provided a simple formula for a particular choice of $M_n$, and we will use Hodges' formula as the definition of $M_n$ in this paper.\footnote{More accurately, we use a special case of Hodges' formula where we fix certain arbitrary choices, which we write explicitly in \eqref{eq:HodgesDet}. We also set $\kappa = \sqrt{32 \pi G} $. See Appendix \ref{app:splitting} for conventions.}   

Our main result is that, for $n > 3$, Hodges' MHV formula for the stripped amplitude $M_n$ satisfies a simple recursion relation that allows it to be computed from $M_{n-1}$:
\begin{equation}\label{eq:result}
    M_n = -\frac{\kappa}{2}\sum_{i = 1}^{n-1} \frac{[n i]}{\langle n i \rangle} \frac{\langle \alpha i \rangle^2}{\langle \alpha n \rangle^2} M_{n-1}( \, \ket{1}, |1], \, \cdots, | i \rangle, |i]{+}\frac{\langle \alpha n\rangle}{ \langle \alpha i \rangle}|n], \, \cdots, \ket{n{-}1}, |n{-}1] \, ).
\end{equation}
Here $\ket{\alpha}$ is an arbitrary reference spinor that appears in Hodges' formula. We use the usual representative $M_3 =  \frac{i \kappa}{2}\left(\frac{\lr{12}^4}{ \lr{1 2} \lr{2 3} \lr{3 1} } \right)^2$ as the base of the recursion, and this generates all $M_{n>3}$ using \eqref{eq:result}.

Interestingly, the above equation has not appeared before in the literature. Its most notable feature is that it sums over all $n-1$ particles in $M_{n-1}$.  The existing well-known recursion relations for the full MHV amplitude $\mathcal{M}_n$, such as (9) in \cite{Hodges:2012ym}, or (2) in \cite{Boucher-Veronneau:2011rwd}, are based on BCFW recursion \cite{Britto:2005fq} and explicitly conserve momentum at each recursive step. See also \cite{Krasnov:2013wsa,Delfino:2014xea,Berends:1988zp,Cachazo:2005ca,Elvang:2007sg,Bianchi:2008pu,Arkani-Hamed:2008owk,Spradlin:2008bu,Trnka:2020dxl,Hu:2022bpa,Paranjape:2023qsq,Koefler:2024pzv}. As a result, these formulas shift more than one particle at a time and always involve a sum over $n-2$ (or fewer) particles.  However, any recursion relation that shifts more than one particle at a time does not act locally on the celestial sphere, so our formula is  more natural from a celestial perspective.

\subsection{Main Result Interpreted as an $\Lw$ Ward Identity}

Our ansatz for the form of  \eqref{eq:result} was motivated by considerations from the celestial holography program. In order to explain these considerations, it is convenient (although strictly speaking not necessary) to parameterize null momenta as
\begin{equation}\label{momentum}
    p^\mu_i = \frac{\omega_i}{2}(1 + z_i \bz_i, \, z_i + \bz_i, \, -i(z_i - \bz_i), \, 1 - z_i \bz_i ),
\end{equation}
their spinor decompositions as 
\begin{equation}
    \hspace{0.5 cm} p_i^{A \dot A} = \begin{pmatrix} p^0_i +  p^3_i & p^1_i - i p^2_i \\ p^1_i + i p^2_i & p^0_i - p^3_i \end{pmatrix} = \lambda_i^A \tlambda_i^{\dot A},
\end{equation}
and the spinors as
\begin{equation}\label{spinorchoices}
   \lambda_i^A = \begin{pmatrix}
        1 \\ z_i
    \end{pmatrix}, \hspace{1 cm} \tlambda_i^{\dot A} = \begin{pmatrix} \omega_i \\ \omega_i \bz_i \end{pmatrix}, \hspace{1 cm} \alpha^A = \begin{pmatrix}
        0 \\ 1
    \end{pmatrix}.
\end{equation}
With these choices, we have 
\begin{equation} \label{eq:choiceresult}
    \lr{ i j} = z_{ij}, \hspace{1 cm} [ ij ] = \omega_i \omega_j \bz_{ij}, \hspace{1 cm} \langle \alpha i \rangle = 1,
\end{equation}
where $z_{ij} = z_i - z_j$.

Within a postulated 2D celestial CFT, we define the operators $G^\pm(z_i, \tlambda_i)$ to correspond to insertions in the  bulk four-dimensional (4D) $S$-matrix of $\pm$ helicity gravitons with momentum $p_i$. If $\omega_i$ is negative (positive) the graviton is incoming (outgoing). 

For tree-level minimally-coupled gravity amplitudes, there is a simple formula for the \textit{holomorphic collinear splitting function} of two gravitons $G^+(z_i, \tlambda_i) G^\pm(z_j, \tlambda_j)$, where $z_{ij} \to 0$ as $\tlambda_i$, $\tlambda_j$ are held fixed. In the language of celestial CFT, this splitting function reads \cite{Guevara:2022qnm,Ren:2023trv} 
\begin{equation} \label{eq:tildesplit}
    \lim_{z_{ij} \to 0} G^+(z_i, \tlambda_i) \, G^\pm(z_j, \tlambda_j) = -\frac{\kappa}{2}\frac{[ij]}{z_{ij}} G^\pm(z_j, \tlambda_j + \tlambda_i),
\end{equation}
which in the 2D boundary theory should be interpreted as an operator product expansion (OPE) of a level-zero $\myw_{1+\infty}$ current algebra, also known as the $\Lw$ algebra. Note that here the $\tilde{\lambda}_i$ are interpreted as color indices for the $\myw_{1+\infty}$ symmetry. (The other common celestial CFT presentation of the splitting functions in the Mellin basis \cite{Pate:2019lpp,Guevara:2021abz} is simply related to \eqref{eq:tildesplit} by a change of basis for the generators of the $\myw_{1+\infty}$ algebra.) We have provided a review of the $\Lw$ algebra, its connection to the splitting function, and its presentation in different bases in Appendix \ref{app:splitting}.

We now explain how the $\myw_{1+\infty}$ OPE is related to our main result  \eqref{eq:result}. Using the spinor choices \eqref{spinorchoices}, equation \eqref{eq:result} says that Hodges' formula for the stripped amplitude $M_n$ can be generated from a 2D celestial correlator that obeys the following two rules. The first rule is that the $n=3$ base case must satisfy
\begin{equation}\label{eq:M3}
    \big\langle G^-(z_1, \tlambda_1) \, G^-(z_2, \tlambda_2)\, G^+(z_3, \tlambda_3) \big\rangle = \frac{i\kappa}{2}\frac{z_{12}^8 }{z_{12}^2 z_{23}^2 z_{31}^2},
\end{equation}
which is the usual expression for $M_3$. The second rule is that the $n$-particle correlator must be determined by the $(n-1)$-particle correlator using the $\Lw$ Ward identity 
\begin{equation}
\begin{split}
\begin{aligned}
    &\big\langle  G^-(z_1, \tlambda_1) \, G^-(z_2, \tlambda_2) \ldots  G^+(z_n, \tlambda_n)  \big\rangle \\
    &\hspace{0.1 cm} = - \frac{\kappa}{2} \sum_{i = 1}^{n-1}  \frac{[ni]}{z_{ni}} \big\langle G^-(z_1, \tlambda_1)\, G^-(z_2, \tlambda_2) \ldots G^+(z_{i} , \tlambda_{i} + \tlambda_n) \ldots G^+(z_{n-1}, \tlambda_{n-1}) \big\rangle.
\end{aligned}
\end{split}
\end{equation} 
From \eqref{eq:result}, we see that these two rules automatically imply that
\begin{equation}
    \big\langle G^-(z_1, \tlambda_1)\, G^-(z_2, \tlambda_2) \ldots G^+(z_n, \tlambda_n) \big\rangle = M_n,
\end{equation}
and therefore the $\Lw$ Ward identity generates Hodges' formula.

The rest of this paper is dedicated to the proof that \eqref{eq:result} holds for any choice of the spinors $\lambda_i$, $\tlambda_i$, $\alpha$, not just the convenient ones in \eqref{spinorchoices}. The proof will employ the matrix-tree theorem \cite{Feng:2012sy},\footnote{The matrix-tree theorem was also recently used to understand the MHV double copy in \cite{Adamo:2024hme}.} and is most directly inspired by recent results on perturbiner expansions in self-dual gravity and the relation of $\Lw$ symmetry to marked tree expansions \cite{Miller:2024oza,Miller:2025wpq} (see in particular the proof of Theorem A.1 in \cite{Miller:2024oza}).

\section{Hodges' Formula, Forests, and Recursion} \label{sec:proof}

In this section, we prove \eqref{eq:result} using Hodges' formula \cite{Hodges:2012ym} and the matrix-tree theorem \cite{Feng:2012sy}. The relevance of trees to the MHV amplitude was noted by NSVW in \cite{Nguyen:2009jk} and anticipated in \cite{Bern:1998sv}. We review the necessary ingredients in Subsection \ref{sec:HFF} and then provide the proof in Subsection \ref{sec:proofdetails}.

\subsection{Hodges' Formula Expressed Using Forests} \label{sec:HFF}

Consider the $n \times n$ matrix $\Psi$ with entries given by  
\begin{equation} \label{eq:Psi}
    \Psi_{ij} = \begin{cases}   - \frac{[ij]}{\langle i j \rangle} a_i a_j & i \neq j \\ \sum^{k=n}_{k=1,k \neq i}   \frac{[ik]}{\langle i k \rangle} a_i a_k & i = j\end{cases},
\end{equation} 
where
\begin{equation}
        a_i = \langle \alpha i \rangle^2  
\end{equation}
for an arbitrary reference spinor $\ket{\alpha}$.

Hodges \cite{Hodges:2012ym} showed that the tree-level graviton MHV amplitude can be written using the determinant of an $(n-3)\times(n-3)$ minor of $\Psi$. In particular,\footnote{In fact, Hodges \cite{Hodges:2012ym} proved a more general form in which any three rows and any three columns can be removed and are not necessarily equal, and $a_i = \langle \alpha i \rangle \langle \beta i \rangle$ with $\ket{\alpha}$ and $\ket{\beta}$ separate arbitrary reference spinors. $\mathcal{M}_n$ is independent of $\ket{\alpha}$, $\ket{\beta}$ and $S_n$-permutation symmetric after momentum conservation is imposed \cite{Hodges:2012ym}. We specialize to removing rows and columns 1,2,3 because our recursion will be applied to the base case $n=3$, and we also set $\alpha = \beta$. We further note that the NSVW tree formula \cite{Nguyen:2009jk} corresponds to the choice $\ket{\alpha} = \ket{1}$, $\ket{\beta} = \ket{2}$ \cite{Feng:2012sy}. } 
\begin{equation} \label{eq:HodgesDet}
    M_n = -i\left(-\frac{\kappa}{2}\right)^{n-2}\lr{12}^8 \, c_{123}^2 \frac{(a_1 a_2 a_3)^2}{(\prod_{\ell = 1}^n a_\ell)^2} | \Psi|^{123}_{123},
\end{equation}
where $| \Psi|^{123}_{123}$ denotes the determinant of the $(n-3)\times(n-3)$ minor of $\Psi$ with rows and columns $1,2,3$ removed, and 
\begin{equation}
    c_{123} = \frac{1}{\langle 12 \rangle \langle 23 \rangle \langle 31 \rangle}.
\end{equation} 
To prove \eqref{eq:result}, we use the matrix-tree theorem to rewrite $M_n$, following Feng and He \cite{Feng:2012sy}. This theorem implies that for the matrix $\Psi$ defined in \eqref{eq:Psi},
\begin{equation} \label{eq:MT}
    |\Psi|^{123}_{123} = \sum_{F \in \mathcal{F}_{123}^n } \left( \prod_{e_{ij} \, \in \, \text{edges}(F)} -\Psi_{ij} \right),
\end{equation} 
where $\mathcal{F}_{123}^n$ is the set of \textit{rooted forests} consisting of three \textit{rooted trees} with roots $1,2,3$ and nodes taking values in $1,\ldots, n$. A rooted tree with root $k$ is a tree containing the node $k$, so the rooted forests have three disconnected trees with the roots $1,2,3$ each in a different tree (i.e. the nodes $1,2,3$ always appear in disconnected trees).

\begin{figure}[H] 
    \centering
    \includegraphics{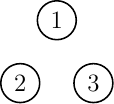}
    \caption{The single rooted forest with 3 trees in $\mathcal{F}_{123}^{3}$, i.e. with $n=3$.}
    \label{fig:3pt}
\end{figure}

\begin{figure}[H] 
    \centering
    \includegraphics{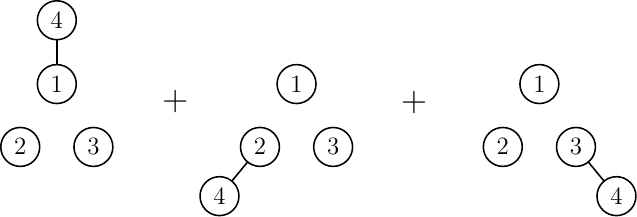}
    \caption{The three rooted forests with 3 trees in $\mathcal{F}_{123}^{4}$, i.e. with $n=4$.}
    \label{fig:4pt}
\end{figure}

\begin{figure}[H]
    \centering
    \includegraphics{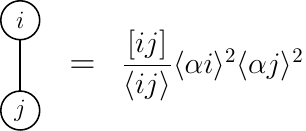}
    \caption{In each graph, the edges correspond to a multiplicative factor $\frac{[ij]}{\langle i j\rangle} \langle \alpha i \rangle \langle \alpha j \rangle.$}
     \label{fig:edgeweight}
\end{figure}

Using \eqref{eq:HodgesDet} and \eqref{eq:MT} we arrive at the forest formula
\begin{equation} \label{eq:HFF}
\begin{aligned}
    M_n &= -i\left(-\frac{\kappa}{2}\right)^{n-2}\lr{12}^8 \left( c_{123} \frac{a_{1} a_{2} a_{3}}{\prod_{\ell = 1}^n a_\ell} \right)^2 \sum_{F \in \mathcal{F}_{123}^n} \left( \prod_{e_{ij} \, \in \, \text{edges}(F)} \frac{[ij]}{\langle i j \rangle} a_i a_j \right) .
\end{aligned}
\end{equation}
For concreteness, we now exhibit low-point examples at $n=3$ and $n=4$. With $n=3$, corresponding to 
\begin{equation}
    M_3 = \frac{i\kappa}{2}\left(\frac{\lr{12}^4}{ \lr{1 2} \lr{2 3} \lr{3 1} } \right)^2, 
\end{equation}
the only possible rooted forest of three trees is simply the three roots themselves, as illustrated in Figure \ref{fig:3pt}.  This will be the base of our recursion. 
When $n=4$, we have
\begin{equation}
    M_4 = \frac{-i\kappa^2}{4}\frac{\lr{12}^6 }{ \lr{2 3}^2 \lr{3 1}^2 } \frac{1}{\lr{\alpha 4}^4} \left( \frac{[14]}{\lr{14}} \lr{\alpha 1}^2 \lr{\alpha 4}^2 + \frac{[24]}{\lr{24}} \lr{\alpha 2}^2 \lr{\alpha 4}^2  + \frac{[34]}{\lr{34}} \lr{\alpha 3}^2 \lr{\alpha 4}^2  \right),
\end{equation}
which is represented by the sum over forests in Figure \ref{fig:4pt}. In each forest, the edges carry the weights shown in Figure \ref{fig:edgeweight}. 

\begin{figure}[H]
    \centering
    \includegraphics{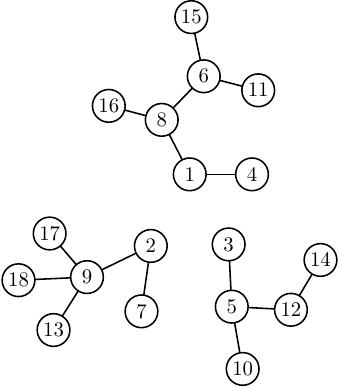}
    \caption{A forest with 3 trees in $\mathcal{F}_{123}^{18}$, i.e. with $n=18$.}
    \label{fig:Fex}
\end{figure}

For a higher-point example, we draw an element of $\mathcal{F}_{123}^{18}$ in Figure \ref{fig:Fex} that would contribute to $M_{18}$.

\subsection{Proof of Recursive Formula} \label{sec:proofdetails}

Having reviewed Feng and He's forest formula \eqref{eq:HFF}, we now use it to prove \eqref{eq:result}.  \\

\begin{figure}[H]
    \centering
    \includegraphics{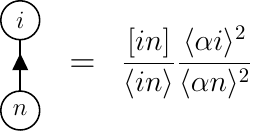}
    \caption{A special arrow edge between nodes $i$ and $n$ corresponding to the overall multiplicative factor for each $i$ in \eqref{eq:result}.}
    \label{fig:inedge}
\end{figure}

\begin{figure}[H]
    \centering
    
    \includegraphics{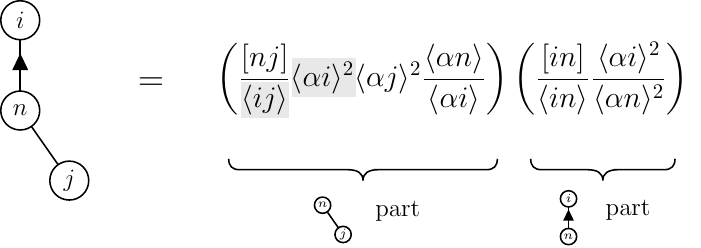}
    \caption{The special rule for edges $e_{jn}$ generated by the recursive formula \eqref{eq:result} that connects to an arrowed $e_{in}$ edge. The highlighted factors emphasize that the weight of $e_{jn}$ depends on $i$, indicated by the directed arrow edge $e_{in}$.} 
    \label{fig:jnedge}
\end{figure}

\begin{figure}[H]
    \centering
    \includegraphics{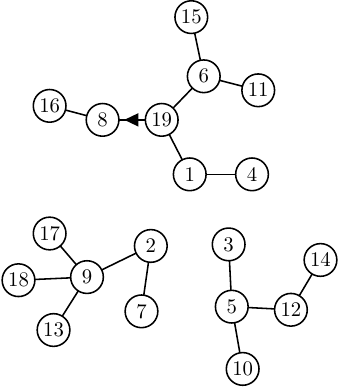}
    \caption{An arrowed rooted forest with three trees in $\mathcal{F}^{19 \to 8}_{123}$, i.e. with $n=19$ and $i =8$. In the recursive formula, this arrowed rooted forest would be generated from the regular rooted forest shown in Figure \ref{fig:Fex}.}
    \label{fig:Farrowex}
\end{figure}

\begin{figure}[H]
    \centering
    \includegraphics{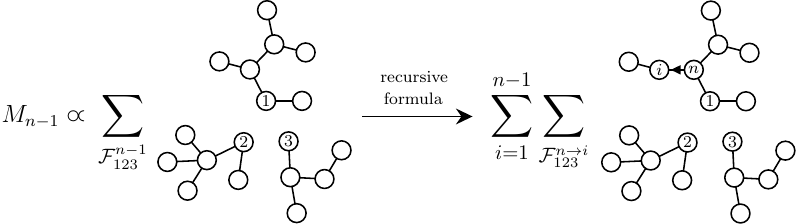}
    \caption{A schematic representation of how the recursive formula \eqref{eq:result} acts on the Hodges forest formula: it takes a sum over forests in $\mathcal{F}^{n-1}_{123}$ to a sum over forests in $\mathcal{F}^{n \to i}_{123}$.}
    \label{fig:graphsthatlooklikethis}
\end{figure}

The recursive step of \eqref{eq:result} involves two modifications of $M_{n-1}$ for each $i$, where $i$ is summed from $1$ to $n-1$. The first is simply multiplication by an overall factor of $\frac{[in]}{\langle i n \rangle} \frac{\langle \alpha i \rangle^2}{\langle \alpha n \rangle^2} $, which we can represent in the forest formula by adding a special edge $e_{in}$ between $i$ and the new node $n$. This edge is drawn with an arrow pointing from $n$ to $i$, with the special weight $\frac{[in]}{\langle i n \rangle} \frac{\langle \alpha i \rangle^2}{\langle \alpha n \rangle^2}$, as shown in Figure \ref{fig:inedge}. (We draw the directed arrow on this edge to anticipate that nodes connected to $n$ will have another special weight, which we will explain shortly.)
The second modification of $M_{n-1}$ is the transformation of the factor, for every forest $F$, of 
\begin{equation}\label{eqrecursenodes}
\begin{aligned}
 \prod_{j \, \in \, \text{neighbors of } i} \frac{[ij]}{\langle i j \rangle} a_i a_j \ \ \ \to  \prod_{j \, \in \, \text{neighbors of } i} \left( \frac{[ij]}{\langle i j \rangle} a_i a_j + \frac{[nj]}{\langle i j \rangle}\frac{\langle \alpha n \rangle}{\langle \alpha i \rangle} a_i a_j \right). 
\end{aligned}
\end{equation}
When expanded, this can be represented in the forest formula by a sum over all combinations of trees where each node $j$ formerly connected to $i$ is instead either 1) still connected to $i$ with an edge $e_{ij}$ weighted by the usual factor in Figure \ref{fig:edgeweight}, or 2) connected to $n$ with an edge $e_{jn}$ weighted by another special factor $\frac{[nj]}{\langle i j \rangle}\frac{\langle \alpha n \rangle}{\langle \alpha i \rangle} a_i a_j$ that depends on $i$. Thus, the weights of edges that are connected to $n$ depend on the fact that $n$ has a directed arrow edge to $i$. The product of special edges $e_{in}$ and $e_{jn}$ is shown in Figure \ref{fig:jnedge}. The highlighted factors in Figure \ref{fig:jnedge} emphasize that the weight of $e_{jn}$ depends on $i$ because of the directed arrow edge $e_{in}$. 
\begin{figure}[H]
    \centering
    \includegraphics{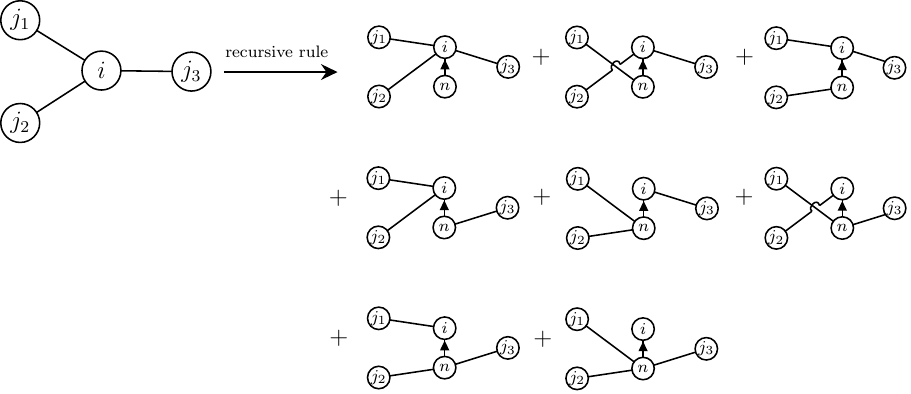}
    \caption{If a collection of $s$ neighbors attach to the node $i$ in a graph, the recursive rule \eqref{eq:result} transforms this into a sum $2^s$ of graphs where $n$ attaches to $i$ by an arrowed edge and all $s$ neighbors can attach to either $i$ or $n$. See \eqref{eqrecursenodes}. In this example $s=3$.}
    \label{fig:recursiverule}
\end{figure}
The action for each $i$ of the recursive formula is illustrated in Figure \ref{fig:recursiverule}. The full result of the recursive formula, summed over $i$, thus transforms the sum over all rooted forests $\mathcal{F}_{123}^{n-1}$ with $n-1$ nodes into a sum over $i$ of all \textit{arrowed rooted forests} with $n$ nodes, i.e.\!\! rooted forests with $n$ nodes in which the node $n$ is special and points to exactly one neighbor node $i$ with an arrowed edge.  We denote the latter set by $\mathcal{F}_{123}^{n\to i}$. An example of a full arrowed rooted forest in $\mathcal{F}^{19 \to 8}_{123}$ is given in Figure \ref{fig:Farrowex}, and a schematic illustration of the full recursive action is shown in Figure \ref{fig:graphsthatlooklikethis}.

\begin{figure}[H]
    \centering
    \includegraphics{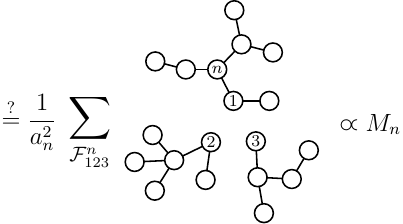}
    \caption{We want to prove that the right-hand side from Figure \ref{fig:graphsthatlooklikethis} is equal to $M_n$, which is a sum over graphs in $\mathcal{F}^n_{123}$.}
    \label{fig:graphsthatlooklikethis2}
\end{figure}

Using the notation of arrowed rooted forests, we now have
\begin{equation} \label{eq:Mnm1}
\begin{aligned}
    &-\frac{\kappa}{2}\sum_{i = 1}^{n-1}\frac{[n i]}{\langle n i \rangle} \frac{\langle \alpha i \rangle^2}{\langle \alpha n \rangle^2} M_{n-1}( \ldots, | i \rangle, |i] + \frac{\langle \alpha n\rangle}{ \langle \alpha i \rangle}|n], \ldots) \\
    &\hspace{0.5 cm} = -i \left( -\frac{\kappa}{2}\right)^{n-2}\lr{12}^8 \sum_{i = 1}^{n-1} \left( c_{123} \frac{a_{1} a_{2} a_{3}}{\prod_{\ell = 1}^{n-1} a_\ell} \right)^2 \sum_{F \in \mathcal{F}_{123}^{n\to i}} \left( \prod_{e_{kl} \, \in \, \text{edges}(F)} w_{kl} \right),  \\
\end{aligned}
\end{equation}
where, assuming without loss of generality that $k < l$, the weights $w_{kl}$ are defined by the edges we described above:
\begin{equation} \label{eq:weightdef}
w_{kl} = \begin{cases} \frac{[kl]}{\langle kl \rangle} a_k a_l,  &k,l \neq n, \\
 \frac{[in]}{\langle i n \rangle} \frac{\langle \alpha i \rangle^2}{\langle \alpha n \rangle^2},  &k= i, l = n,  \\
 \frac{[nk]}{\langle i k \rangle}\frac{\langle \alpha n \rangle}{\langle \alpha i \rangle} a_i a_k, &k \neq i, l = n. \\
\end{cases}
\end{equation}
Comparing \eqref{eq:Mnm1} and \eqref{eq:HFF}, we see that we can prove that the recursive formula holds if we show that the sum over $i \in [1,n-1]$ of arrowed rooted forests $\mathcal{F}_{123}^{n\to i}$ appearing in \eqref{eq:Mnm1} is a factor of $\frac{1}{a_{n}^2}$ times the sum of regular  rooted forests $\mathcal{F}_{123}^{n}$ with regular edges (i.e. where we forget the decoration of the arrow). In pictures, we want to prove that the right-hand side of Figure \ref{fig:graphsthatlooklikethis} is equal to Figure \ref{fig:graphsthatlooklikethis2}. 

\begin{figure}[H]
    \centering
    \includegraphics{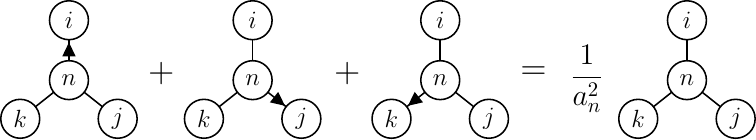}
    \caption{Summing the placement of the arrowed edge over the neighbors of $n$, labeled $\{i,j,k\}$, is equal to a graph with the arrow removed multiplied by $1/a_n^2$.}
    \label{fig:rootedsum}
\end{figure}

To show that \eqref{eq:Mnm1} and \eqref{eq:HFF} are equal, first consider a particular arrowed rooted forest $F^{n\to i,\underline{m}} \in \mathcal{F}_{123}^{n\to i}$ in which the node $n$ is connected to the specific set of neighbors $\underline{m}$, where $\underline{m}$ is a subset of $\{1,\cdots,n-1\}$ of size $m$.  In particular, $\underline{m}$ includes the neighbor $i$, to which $n$ is connected by an arrow edge. Now define another arrowed rooted forest $F^{n\to j,\underline{m}}$ which -- except for the fact that the arrowed edge points to $j$ instead of $i$ -- has the identical structure as $F^{n\to i,\underline{m}}$, with $n$ attached to the identical set of neighbors $\underline{m}$.
Indexing the neighbors of $n$, i.e. the set $\underline{m}$, by $\iota$, we can consider the terms in \eqref{eq:Mnm1} that sum $F^{n \to \iota,\underline{m}}$ over all $\iota \in \underline{m}$.  We will now show that the sum of these terms is exactly equal to $\frac{1}{a_n^2}$ times the corresponding regular rooted forest (i.e. where we forget the decoration of the arrow), which we denote  $F^{n,\underline{m}} \in \mathcal{F}_{123}^n$. Then, our result \eqref{eq:result} will follow. Figure \ref{fig:rootedsum} shows an example with $\underline{m} = \{i,j,k\}$. 

The sum over $\iota \in \underline{m}$ of the rooted forests $F^{n\to \iota, \underline{m}}$ explicitly becomes a sum $\sum_{\iota \in \underline{m}} \prod_{\iota \in \underline{m}} w_{\iota n}$ over the neighbors of $n$, multiplied by an overall factor for the rest of the edges in the arrowed rooted forests. This overall factor is already identical to the corresponding overall factor in the regular (non-arrowed) rooted forest $F^{n,\underline{m}}$. A few lines of algebra, presented in Appendix \ref{app:proofdetails}, show that 
\begin{equation} \label{eq:iotasum}
\sum_{\iota \in \underline{m}}  \left( \prod_{\iota \in \underline{m}} w_{\iota n} \right) = \frac{1}{a_n^2} \left( \prod_{\iota \in \underline{m}} \frac{[\iota n]}{\langle \iota n \rangle} a_\iota a_n \right),
\end{equation}
i.e., the sum over $\iota \in \underline{m}$ reproduces the product over the neighbors of $n$ with the usual weights in the corresponding unarrowed forest $F^{n,\underline{m}}$, multiplied by an additional factor of $\frac{1}{a_n^2}$.

It is clear that the recursive formula \eqref{eq:Mnm1} is precisely  a sum over all such possible sets of arrowed rooted forests, multiplied by the overall factor $\left( c_{123} \frac{a_{1} a_{2} a_{3}}{\prod_{\ell = 1}^{n-1} a_\ell} \right)^2$. We have demonstrated that this is exactly the sum over the set $\mathcal{F}_{123}^{n}$ of unarrowed forests multiplied by the overall factor $\left( c_{123} \frac{a_{1} a_{2} a_{3}}{\prod_{\ell = 1}^{n-1} a_\ell} \right)^2 \frac{1}{a_n^2} = \left( c_{123} \frac{a_{1} a_{2} a_{3}}{\prod_{\ell = 1}^{n} a_\ell} \right)^2$, which is what we wanted to show. Thus we have proved \eqref{eq:result}.

\section{Discussion}\label{sec:discussion}

In this paper, we have proven that the recursive formula \eqref{eq:result} generates Hodges' formula for the stripped amplitude \eqref{eq:HodgesDet}. We used the matrix-tree theorem as an intermediate step in the proof. We will now make a few remarks about this result. 

Our celestial ``dictionary'' for computing the MHV amplitude $\mathcal{M}_n$ instructs us to start with the base case for $M_3$, repeatedly apply the recursive Ward identity to compute $M_n$, and then simply multiply the whole expression by the momentum conserving delta function at the end, as in \eqref{eq:strippedamplitude}. Importantly, while the stripped amplitude $M_n$ is not invariant under the choice of the reference spinor $\ket{\alpha}$, Hodges showed that the full amplitude $\mathcal{M}_n$ will be invariant when momentum is conserved.

We also note that without momentum conservation, \eqref{eq:HodgesDet} only makes manifest the $S_3 \times S_{n-3}$ subgroup of full permutation symmetry.  Hodges showed that $\mathcal{M}_n$ is fully $S_n$ permutation symmetric when momentum conservation is finally imposed, modulo the factor of $\lr{12}^8$ which can be understood as coming from supermomentum conservation \cite{Grisaru:1976vm,Grisaru:1977px}. The fact that we are considering a recursive formula in which (super)momentum is naturally stripped is highly reminiscent of amplitudes in momentum-twistor space and suggests a plausible connection back to the original argument of Hodges. 

The violation of momentum conservation by our one-particle recursion is a reflection of the simple fact that away from the strict holomorphic collinear limit, two massless particles with momenta $p_1^2=0$ and $p_2^2=0$ have $(p_1 + p_2)^2 \neq 0$. In the language of celestial OPEs, this means that although the leading term in OPEs derived from holomorphic collinear limits is covariant under a chiral half of Poincar\'e \cite{Stieberger:2018onx,Himwich:2021dau} involving two translations and three Lorentz transformations, it is \textit{not} covariant on its own under the other two translations, as emphasized recently in \cite{Kulp:2024scx}. This is connected to the fact that one chiral half of Poincar\'e is contained within a $\myw_{1+\infty} \subset \Lw$ subalgebra, while the two other translations require going one rung down in the loop parameter. Full Poincar\'e covariance is expected to become manifest once further subleading terms are included properly in the OPE.

For the reasons discussed above, it is actually possible to generate exactly two of the four components of the momentum-conserving delta function using our $\Lw$ construction. We briefly outline this now. With the spinor choices \eqref{spinorchoices}, in which  $(\tlambda_i^{\dot 1},\tlambda_i^{\dot 2}) = \left(p^0_i + p^3_i, p^1_i - i p^2_i\right)$, the delta function becomes $\delta^2( \sum_i \tlambda_i) \delta^2( \sum_i z_i \tlambda_i)$. The two components $\delta^2( \sum_i \tlambda_i)$ are generated by the recursive rule \eqref{eq:result} if one simply multiplies the $n=3$ base case \eqref{eq:M3} by $\delta^2(\tlambda_1 + \tlambda_2 + \tlambda_3)$. This fact was also noted by \cite{Guevara:2022qnm,Guevara:2024vlc} and is natural because $\delta^2(\tlambda_1 + \tlambda_2)$ is the invariant Killing form of the $\myw_{1+\infty}$ algebra. 

However, it is not apparent how to generate the other two components of the delta function $\delta^2( \sum_i z_i \tlambda_i)$ in a similarly natural way. Indeed,  \cite{Guevara:2022qnm} discussed the same recursion relation applied to $\mathcal{M}_{n-1}$ instead of $M_{n-1}$, and found that it generated $\mathcal{M}_n$ as well as additional terms (needed to conserve momentum) corresponding to poles at infinity. See for instance equations (33)-(35) of \cite{Guevara:2022qnm}. See also \cite{Ren:2023trv}.  We hope to understand the proper interpretation of these terms in future work. Alternatively all four components of the delta function could be constructed with a celestial leaf correlator, as in \cite{Melton:2024akx,Melton:2023bjw,Mol:2024etg}.

We additionally comment that the right-hand side of \eqref{eq:result} is precisely the sum of exponentiated soft factors that organize the universal part of the so-called tower of soft theorems \cite{Hamada:2018vrw,Li:2018gnc,Guevara:2019ypd,Bautista:2019tdr,Adamo:2021lrv,Guevara:2022qnm}.\footnote{Recursion involving the exponentiated soft factor, accompanied by the shift of a second particle to use the inverse soft recursion formula, was also studied in a celestial context in \cite{Ren:2023trv,Guevara:2019ypd,Guevara:2022qnm}.} This universal part generates the action of $\myw_{1+\infty}$ on hard massless particles \cite{Himwich:2021dau,Adamo:2021lrv}. However, because it does not conserve momentum, as it currently stands it cannot exactly specify \textit{any} (momentum-conserving) amplitude. Instead, the most we can expect is that it generates a stripped amplitude, exactly as we have shown. We hope to return to the question of momentum conservation in future work. 

As a final note, the reader may also wonder if, besides Hodges' formula, there are other stripped MHV amplitude representatives that satisfy the recursive formula \eqref{eq:result}. We know of exactly one other: the somewhat mysterious formula (5.54) of \cite{Miller:2024oza} with the reference spinors $\ket{\iota}$, $\ket{o}$ replaced with $\ket{\alpha}$. This formula expresses $M_n$ as a sum over connected trees (as opposed to forests) containing all the positive and the two negative helicity gravitons, the latter with special edge factors involving two extra antiholomorphic reference spinors $|\xi]$ and $|\chi]$. It turns out that the $n$-point stripped amplitude of \cite{Miller:2024oza} can be generated from its three-point case using \eqref{eq:result}. The proof works along the same lines as the proof given in this paper, although one must account for the special $|\xi]$, $|\chi]$ edges.

\section*{Acknowledgements}

We thank Freddy Cachazo, Monica Pate, David Skinner, Jaroslav Trnka, and Anastasia Volovich for useful conversations. NM gratefully acknowledges support from the Sivian Fund at the Institute for Advanced Study and DOE grant DE-SC0009988, along with the Princeton Gravity Initiative. AG acknowledges support from DOE grant DE-SC0007870. EH acknowledges  support from the Princeton Center for Theoretical Science and the Princeton Gravity Initiative.

\appendix

\section{Algebraic Details of Proof} \label{app:proofdetails}

In this appendix, we prove the equality \eqref{eq:iotasum} of sums over neighbors of $n$ shown in Figure \ref{fig:rootedsum}. The useful identity \eqref{eq:usefulID} used in the proof is derived in Subsection \ref{sec:usefulID}. 

The sum over $\iota \in \underline{m}$ can be indexed by  $\iota = 1, \ldots, m$ (note that this index is only a way to enumerate the neighbors of $n$, and does not reflect the actual particle label of the nodes). Then, using the definition $a_j = \langle \alpha j \rangle^2$ and the weights \eqref{eq:weightdef}, we have 

\begin{equation}
    \begin{aligned}
        \sum_{\iota \in \underline{m}}  \left( \prod_{\iota \in \underline{m}} w_{\iota n} \right) &=  \sum_{\iota =1}^m \frac{[ \iota n]}{\langle \iota n \rangle} \frac{\langle \alpha \iota \rangle^2}{\langle \alpha n \rangle^2} \prod_{\substack{j = 1\\ j \neq \iota}}^m \frac{[n j]}{\lr{ \iota j} } \lr{\alpha \iota}^2 \lr{\alpha j}^2 \frac{\lr{\alpha n}}{\lr{\alpha \iota}} \\
        &= \sum_{\iota = 1}^m \frac{[\iota n]}{\lr{\iota n}} \lr{\alpha \iota}^{2+m-1} \lr{\alpha n}^{-2+m-1} \prod_{\substack{j = 1\\ j \neq \iota}}^m \frac{[n j]}{\lr{\iota j}} \lr{\alpha j}^2 \\
        &= - \left( \prod_{k =1}^m [ n k] \lr{\alpha k}^2 \right) \lr{\alpha n}^{2m - 4} \sum_{\iota = 1}^m \left( \frac{\lr{\alpha \iota}}{\lr{\alpha n}}\right)^{m-1} \frac{1}{\lr{\iota n}}  \prod_{\substack{j = 1\\ j \neq \iota}}^m \frac{1}{ \lr{\iota j}} .
        \end{aligned}
        \end{equation}
        Now, using the identity \eqref{eq:usefulID}, we can rewrite the sum over $\iota$ as
        \begin{equation}
        \begin{aligned}
        \sum_{\iota \in \underline{m}}  \left( \prod_{\iota \in \underline{m}} w_{\iota n} \right) &= \left( \prod_{k =1}^m [n k] \lr{\alpha k}^2 \right) \lr{\alpha n}^{2m -4} \prod_{\iota = 1}^m \frac{1}{\lr{n \iota}} \\
        &= \frac{1}{\lr{\alpha n}^4}\prod_{\iota = 1}^m \frac{[n \iota]}{\lr{n \iota}}\lr{\alpha \iota}^2 \lr{\alpha n}^2 \\
        &= \frac{1}{a_n^2} \left( \prod_{\iota \in \underline{m}} \frac{[\iota n]}{\langle \iota n \rangle} a_\iota a_n \right),
    \end{aligned}
\end{equation}
and thus we have shown \eqref{eq:iotasum}. 

\subsection{A Useful Identity} \label{sec:usefulID}

Here, we prove the useful ``partial fractions'' identity
\begin{equation} \label{eq:usefulID}
    \sum_{\iota = 1}^m \left( \frac{\langle \alpha \iota \rangle}{\langle \alpha n \rangle} \right)^{m-1} \frac{1}{\langle \iota n \rangle} \prod_{\substack{j = 1\\ j \neq \iota}}^m \frac{1}{\langle  \iota j \rangle} = - \prod_{\iota = 1}^m \frac{1}{\langle n \iota \rangle}
\end{equation}
by induction. To check that the equation is plausibly true, notice that if we treat the spinor helicity variable $\ket{n}$ as a variable in $\mathbb{C}P^1$, then both sides of the expression have all the same poles and residues. We remind the reader that $n \notin \{ 1, \ldots, m\}$.

The case $m=1$ is trivially true. The base case $m=2$ follows from the Schouten identity. Now, we assume \eqref{eq:usefulID} holds for $m$ legs and show that it holds for $m+1$. First, separate the $m+1$ sum into a sum over $m$, plus the $(m+1)$-st term:
\begin{equation}
\begin{aligned}
\sum_{\iota=1}^{m+1} \left( \frac{\langle \alpha \iota \rangle}{\langle \alpha n \rangle}\right)^m &\frac{1}{\langle \iota n \rangle} \prod_{\substack{j = 1\\ j \neq \iota}}^{m+1} \frac{1}{\langle  \iota j \rangle} \\ &= \frac{1}{\langle n (m{+}1) \rangle}\sum_{\iota=1}^{m} \left( \frac{\langle \alpha \iota \rangle}{\langle \alpha n \rangle}\right)^{m-1} \frac{\langle \alpha \iota \rangle}{\langle \alpha n \rangle} \frac{\langle n (m{+}1) \rangle}{\langle \iota (m{+}1) \rangle} \frac{1}{\langle \iota n \rangle} \prod_{\substack{j = 1\\ j \neq \iota}}^{m} \frac{1}{\langle  \iota j \rangle} \\ 
&\qquad + \left( \frac{\langle \alpha (m{+}1) \rangle}{\langle \alpha n \rangle}\right)^{m} \frac{1}{\langle (m{+}1) n \rangle} \prod_{j=1}^{m} \frac{1}{\langle  (m{+}1) j \rangle} .
\end{aligned}
\end{equation}
Now, in the first line, use the Schouten identity
\begin{equation}
\langle \alpha \iota \rangle\langle n (m{+}1) \rangle = \langle \alpha n \rangle\langle \iota (m{+}1) \rangle - \langle \alpha (m{+}1) \rangle\langle \iota n  \rangle
\end{equation}
to find
\begin{equation}
\begin{aligned}
\sum_{\iota=1}^{m+1} \left( \frac{\langle \alpha \iota \rangle}{\langle \alpha n \rangle}\right)^m \frac{1}{\langle \iota n \rangle} \prod_{\substack{j = 1\\ j \neq \iota}}^{m+1} \frac{1}{\langle  \iota j \rangle} &= \frac{1}{\langle n (m{+}1) \rangle}\sum_{\iota=1}^{m} \left( \frac{\langle \alpha \iota \rangle}{\langle \alpha n \rangle}\right)^{m-1}  \frac{1}{\langle \iota n \rangle} \prod_{\substack{j = 1\\ j \neq \iota}}^{m} \frac{1}{\langle  \iota j \rangle} \\ 
& - \frac{1}{\langle n (m{+}1) \rangle}\Bigg[\sum_{\iota=1}^{m} \left( \frac{\langle \alpha \iota \rangle}{\langle \alpha n \rangle}\right)^{m-1} \frac{\langle \alpha (m{+}1)  \rangle}{\langle \alpha n \rangle \langle \iota (m{+}1) \rangle} \prod_{\substack{j = 1\\ j \neq \iota}}^{m} \frac{1}{\langle  \iota j \rangle} \\
&\qquad \qquad \qquad \qquad +\left( \frac{\langle \alpha (m{+}1) \rangle}{\langle \alpha n \rangle}\right)^{m}  \prod_{j=1}^{m} \frac{1}{\langle  (m{+}1) j \rangle}  \Bigg].
\end{aligned}
\end{equation}
The first line involves the sum in the inductive assumption \eqref{eq:usefulID}, as does the term in brackets: note that 
\begin{equation}
\begin{aligned}
\sum_{\iota=1}^{m} \left( \frac{\langle \alpha \iota \rangle}{\langle \alpha n \rangle}  \right)^{m{-}1} &\frac{ \langle \alpha (m{+}1)  \rangle }{\langle \alpha n \rangle \langle \iota (m{+}1) \rangle} 
\prod_{\substack{j = 1\\ j \neq \iota}}^{m} \frac{1}{\langle  \iota j \rangle} \\ &= \left( \frac{\lr{\alpha(m{+}1)}}{\lr{\alpha n}} \right)^m \sum_{\iota=1}^{m} \left( \frac{\lr{\alpha \iota}}{\lr{\alpha(m{+}1)}} \right)^{m-1} \frac{1}{\lr{\iota(m{+}1)}} \prod_{\substack{j = 1\\ j \neq \iota}}^{m} \frac{1}{\langle  \iota j \rangle}  \\
&= - \left( \frac{\lr{\alpha(m{+}1)}}{\lr{\alpha n}}\right)^m \prod_{j=1}^{m} \frac{1}{\langle  (m{+}1) j \rangle}.
\end{aligned}
\end{equation}
Therefore, the term in brackets vanishes, and we have 
\begin{equation}
\begin{aligned}
\sum_{\iota=1}^{m+1} &\left( \frac{\langle \alpha \iota \rangle}{\langle \alpha n \rangle}\right)^m \frac{1}{\langle \iota n \rangle} \prod_{\substack{j = 1\\ j \neq \iota}}^{m+1} \frac{1}{\langle  \iota j \rangle} = \frac{1}{\langle n (m{+}1) \rangle}\left(- \prod_{\iota = 1}^m \frac{1}{\langle n \iota \rangle} \right) = - \prod_{\iota = 1}^{m+1} \frac{1}{\langle n \iota \rangle} 
\end{aligned}
\end{equation}
which is what we wanted to show.

\section{Quick Review of $\Lw$}\label{app:splitting}

In this appendix we briefly review the holomorphic graviton collinear splitting function and its relation to the $\Lw$ algebra, following \cite{Guevara:2021abz,Strominger:2021mtt,Miller:2025wpq}. See also \cite{Monteiro:2011pc}.

Consider a scattering amplitude with two external gravitons of momenta $p_1$ and $p_2$, where the momenta $p_i^{A \dot A} = \lambda_i^A \tlambda_i^{\dot A}$ can be parameterized by $p^\mu_i =  p^\mu(z_i, \tlambda_i)$, with $\lambda_i^A = (1,z_i)$, $\tlambda_i^{\dot A} = (\omega_i, \omega_i \bz_i)$, as in \eqref{momentum} and \eqref{spinorchoices}. We specialize to this choice throughout this appendix. One can show that
\begin{equation}
    p^\mu(z_1, \tlambda_1) + p^\mu(z_2, \tlambda_2) = p^\mu(z_1, \tlambda_1 + \tlambda_2) + \mathcal{O}(z_{12})
\end{equation}
so that in the holomorphic collinear limit $z_{12} \to 0$ with $\tlambda_1$, $\tlambda_2$ fixed, we have $p_1^2 = p_2^2 = 0$ while $(p_1 + p_2)^2 \to 0$.
Defining $P = p_1 + p_2$, in the limit that all three momenta $p_1, p_2$, and $P$ go on-shell, the tree-level collinear splitting function is the product of the standard three-point Feynman vertex  and the  $P$ propagator.\footnote{To quickly obtain these splitting functions, simply use the MHV$_3$ ($\overline{\text{MHV}}_3$)  formula $\frac{i\kappa}{2}\frac{\lr{ab}^6}{\lr{ab}^2 \lr{ac}^2}$  $\left(\frac{i\kappa}{2}\frac{[ab]^6}{[ab]^2 [ac]^2}\right)$ divided by the $P$ propagator $\frac{i}{\lr{12}[12]}$, and use momentum conservation $|1]\langle 1 | + |2]\langle 2 | = |P]\langle P |$ to express these functions in the form we give here. Note that these splitting functions transform covariantly with the proper weight under Lorentz ${\rm SL}(2,\mathbb{C})$.} Using $s_i$ to denote the helicty of particle $i$, we consider the splitting function $\mathrm{Split}^{s_1, s_2}_{s_P}$ with particles 1, 2, and $P$ as illustrated below.\footnote{We use $(+,-,-,-)$ signature, $S_{EH} = -\frac{2}{\kappa^2} \int d^4x \sqrt{-g}~R$, $R_{\mu\nu} = R^{\lambda}_{\  \mu\lambda\nu}$, $\kappa = \sqrt{32 \pi G}$, and $g_{\mu\nu} = \eta_{\mu\nu} + \kappa h_{\mu\nu}$. We also have $p_1 \cdot p_2 = \frac{1}{2}\langle 12 \rangle [12]$ and $\varepsilon_+^{A\dot{A}} = -i\sqrt{2}\frac{\alpha^A\tilde{\lambda}^{\dot{A}}}{\langle \alpha \lambda \rangle}$,  $\varepsilon_-^{A\dot{A}} = -i\sqrt{2}\frac{\lambda^A\tilde{\alpha}^{\dot{A}}}{[ \tilde{\lambda} \tilde{\alpha} ]}$.} It is defined such that
\begin{equation}
    \lim_{|1\rangle \to |2 \rangle} \mathcal{M}^{\rm tree}(1^{s_1}, 2^{s_2}, \ldots) = \sum_{s_P = \pm 2}\mathrm{Split}^{s_1s_2}_{s_P} \mathcal{M}^{\rm tree}(P^{-s_P}, \ldots). 
\end{equation}

If particles 1 and 2 are both positive helicity, the two possible splitting functions are
\begin{equation}
    \begin{matrix}
        \adjincludegraphics[valign=c] {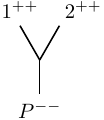}
        \\ \\
        \mathrm{Split}^{+2, +2}_{-2} = -\frac{\kappa}{2}\frac{[1 2]}{\lr{1 2}} \frac{\lr{P \alpha}^4}{\lr{1 \alpha}^2 \lr{2 \alpha}^2}
    \end{matrix}
    \hspace{2 cm}
    \begin{matrix}
        \adjincludegraphics[valign=c] {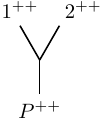} \\
        \\
        \hspace{1.3cm} \mathrm{Split}^{+2, +2}_{+2} = 0\hspace{1.3cm}
    \end{matrix}
\end{equation}
with $\ket{\alpha}$ a reference spinor. Notice that only the $\mathrm{Split}^{+2,+2}_{-2}$ channel contributes. In the holomorphic collinear limit, $\ket{1}$, $ \ket{2}$, and $\ket{P}$ all become equal and we arrive at the celestial OPE
\begin{equation}\label{OpOpope}
    \lim_{z_1 \to z_2} G^+(z_1, \tlambda_1) \, G^+(z_2,\tlambda_2) = -\frac{\kappa}{2}\frac{[12]}{z_{12}} G^+(z_2, \tlambda_1 + \tlambda_2).
\end{equation}
Meanwhile, if particle 2 is negative helicity, the splitting functions are
\begin{equation}
     \begin{matrix}
        \adjincludegraphics[valign=c] {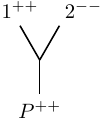}
        \\ \\ 
        \mathrm{Split}^{+2, -2}_{+2} = -\frac{\kappa}{2}\frac{[1 2]}{\lr{1 2}} \frac{\lr{2 \alpha}^6}{\lr{P \alpha}^4 \lr{1 \alpha}^2}
    \end{matrix}
    \hspace{2 cm} 
    \begin{matrix}
        \adjincludegraphics[valign=c] {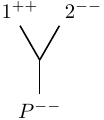}
         \\ \\
        \mathrm{Split}^{+2, -2}_{-2} = -\frac{\kappa}{2}\frac{\lr{1 2}}{[1 2]} \frac{\lr{\alpha P}^4 \lr{\alpha 2}^2}{\lr{\alpha 1}^6 }.
    \end{matrix}
\end{equation}
Only the $\mathrm{Split}^{+2,-2}_{+2}$ channel is singular in the $z_{12} \to 0$ limit, giving us the OPE
\begin{equation} \label{eq:OpOm}
    \lim_{z_1 \to z_2} G^+(z_1, \tlambda_1) \, G^-(z_2,\tlambda_2) = -\frac{\kappa}{2}\frac{[12]}{z_{12}} G^-(z_2, \tlambda_1 + \tlambda_2).
\end{equation}

With these OPEs in hand, we now explain why the $G^+ G^+$ splitting function takes the form of a zero-level Kac-Moody current OPE with a $\myw_{1+\infty}$ color algebra.  $G^-$ can be considered as a charged operator with respect to the current $G^+$.

Abstractly, the $\myw_{1+\infty}$ algebra is simply the algebra of functions of  two variables, say $\mathsf{q}$ and $\mathsf{p}$, where the Lie bracket is the Poisson bracket. This algebra is spanned by a basis of  2D plane waves. Defining the generators and Poisson bracket as
\begin{equation}
    \bT_{\tlambda} \equiv \exp( i \, \mathsf{q} \, \tilde{\lambda}_{\dot{2}} + i \, \mathsf{p} \, \tilde{\lambda}_{\dot{1}} ), \hspace{1 cm} [  f , g  ] \equiv  \partial_{\mathsf{q}} f \, \partial_{\mathsf{p}} g - \partial_{\mathsf{q}} g \, \partial_{\mathsf{p}} f ,
\end{equation}
and using $[ij] = -\tlambda_{i , \dot 1} \tlambda_{j , \dot 2} + \tlambda_{i,\dot 2} \tlambda_{j,\dot 1} $, the commutation relation of $\myw_{1+\infty}$ is
\begin{equation}
    [\bT_{\tlambda_1}, \bT_{\tlambda_2}] = -[12] \bT_{\tlambda_1 + \tlambda_2}.
\end{equation}
The corresponding current algebra matches the OPE \eqref{OpOpope} up to a multiplicative factor of the coupling $\frac{\kappa}{2}$. The $\text{L}$ in $\Lw$ stands for ``loop,'' corresponding to the modes one can take of the holomorphic Kac-Moody currents $G^+(z, \tlambda)$.

In the context of celestial holography, following \cite{Strominger:2021mtt}, it is natural to define  Mellin-transformed and antiholomorphic-mode-expanded operators\footnote{Note that with the conventions \eqref{spinorchoices}, the $G^+$ polarization vectors $\varepsilon_{+}$ are proportional to $\omega$.}
\begin{equation}
\begin{aligned}
    w^{p}_m(z) &= \frac{1}{2} i^{2-2p}(p{+}m{-}1)!(p{-}m{-}1)!  ~\lim_{\epsilon \to 0} \epsilon \oint \frac{d \bz}{2\pi i} \frac{1}{\bz^{p-m}} \int_0^\infty \frac{d \omega}{\omega} \, \omega^{2- 2p + \epsilon} G^+(z,\tilde{\lambda})
    \end{aligned}
    \end{equation}
    for the positive helicity gravitons. We  likewise make the choice
    \begin{equation}
    \begin{aligned}
    O^{p,-}_m(z) &= \frac{1}{2} i^{2-2p}(p{+}m{-}1)!(p{-}m{-}1)!  ~\lim_{\epsilon \to 0} \epsilon \oint \frac{d \bz}{2\pi i} \frac{1}{\bz^{p-m}} \int_0^\infty \frac{d \omega}{\omega} \, \omega^{2- 2p + \epsilon} G^-(z,\tilde{\lambda})
    \end{aligned}
\end{equation}
for the negative helicty gravitons, which we think of as charged operators with respect to the $w^p_m$ currents. These have the OPEs
\begin{equation}
\begin{aligned}
    \lim_{z_{12} \to 0} w^p_m(z_1) w^q_n(z_2) &= -\frac{1}{z_{12}} \frac{\kappa}{2} \left( m(q-1)-n(p-1)\right) w^{p+q-2}_{m+n}(z_2), \\
    \lim_{z_{12} \to 0} w^p_m(z_1) O^{q,-}_n(z_2) &= -\frac{1}{z_{12}} \frac{\kappa}{2} \left( m(q-1)-n(p-1)\right) O^{p+q-2,-}_{m+n}(z_2) .
\end{aligned}
\end{equation}
It should be noted that the $w^p_m$ operators are just the $G^+$ operators employing a different basis of generators for the $\myw_{1+\infty}$ algebra. These new generators, which are just monomials in $\mathsf{q}$, $\mathsf{p}$, are given by
\begin{equation}
    \bT^{p}_{m} \equiv \frac{1}{2} \mathsf{q}^{p+m-1} \mathsf{p}^{p-m-1}, \hspace{1 cm} [\bT^p_m , \bT^q_n ] = \left(m(q-1)-n(p-1)\right) \bT^{p+q-2}_{m+n},
\end{equation}
and can be related to the plane wave basis using the formula 
\begin{equation}
i^{2-2p}(p+m-1)!(p-m-1)! \lim_{\epsilon \to 0} \epsilon \oint \frac{d \bz}{2\pi i} \frac{1}{\bz^{p-m}} \int_0^\infty \frac{d \omega}{\omega} \, \omega^{2- 2p + \epsilon} e^{ i \omega( \mathsf{q} + \mathsf{p} \bz)} = \mathsf{q}^{p+m-1} \mathsf{p}^{p-m-1}.
\end{equation}
When these polynomials contain only positive powers of $\mathsf{q}$ and $\mathsf{p}$, i.e. $p = 1, 0, -1, \ldots$ and $m = 1-p, \ldots, p-1$, we call them elements of the wedge subalgebra $\myw_\wedge \subset \myw_{1+\infty}$.

\bibliography{mhv_bib.bib}
\bibliographystyle{jhep}

\end{document}